\documentclass[titlepage,12pt]{utarticle}
\usepackage{epsf}
\usepackage{amsmath,amsfonts,latexsym}
\usepackage{array}
\usepackage[nosort]{cite}
\long\def\omit#1{}

\numberwithin{equation}{section}

\def\Tr{\text{tr}}
\begin{document}

\preprint{UTTG--05--98\\
HUTP-98/A066\\
{\tt hep-th/9809188}\\
}

\title{The Operator Product Expansion for Wilson Loops and Surfaces 
in the Large $N$ Limit} 

\author{David Berenstein\thanks{Present address: Department of
                Physics, University of Illinois at Urbana-Champaign, 
		Urbana, IL 61801 \hfill\vskip .05em
	\hspace*{.37em}$^\dagger$Research supported in part by
                the Robert A. Welch Foundation and NSF Grant
                PHY~9511632. \hfill\vskip .05em
	\hspace*{.37em}$^\ddagger$Research
        supported in part by grants DE-FG02-96ER40559 and NSF
        PHY~9407194. \hfill\vskip .05em 
	\hspace*{.37em} Email: {\tt berenste@hepux0.hep.uiuc.edu,
        rcorrado@zippy.ph.utexas.edu, \hfill\vskip .05em 
	\hspace*{3.4em} fischler@physics.utexas.edu, 
	malda@pauli.harvard.edu} }\hspace*{6pt}$^{,\dagger}$, 
	Richard Corrado$^\dagger$, Willy Fischler$^\dagger$,} 
	
\oneaddress{Theory Group, Department of Physics\\
        University of Texas at Austin\\
        Austin TX 78712 USA  \\
\vspace*{5mm} 
 {\rm \large and }\\     
\vspace*{5mm} 	
{\rm \large Juan Maldacena $^\ddagger$} 
	\\ {~}\\
	Lyman Laboratory of Physics\\
	Harvard University \\	
	Cambridge, MA 02138 USA  }

\date{September 25, 1998}

\Abstract{The operator product expansion for ``small'' Wilson loops
in $\CN=4$, $d=4$ SYM is  studied.
 The OPE coefficients are
calculated in the large $N$ and $g_{\text{YM}}^2N$ limit by 
exploiting the AdS/CFT  correspondence. 
We also consider Wilson surfaces in the 
$(0,2)$,
$d=6$ superconformal theory. In this case, we find
that the UV divergent terms include 
a term proportional to the rigid string action.
}

\maketitle


\renewcommand{\baselinestretch}{1.25} \normalsize

\section{Introduction}

Over the last year, the connection between 
anti--de Sitter (AdS) spaces and conformal field theories
(CFTs)~\cite{Maldacena:LargeN,Gubser:Correlators,Witten:Holography},
 has provided a method to 
study strongly coupled field theories.  
In gauge theories, a natural observable is the Wilson loop.
In the gravity description, these are related to string worldsheets ending
on the boundary of AdS. 
In this paper, we show how to calculate the operator product expansion
(OPE) of the Wilson loop in which we can approximate a Wilson loop
by local operators when it is small. Since we are dealing with
a conformal field theory, the Wilson loop should be small compared to 
the distances which separate it from other Wilson loops or other
operators. 

Our strategy for computing the OPE of a small Wilson loop in the $4D$
SYM theory is as follows. At large $N$ and 
$g_{\text{YM}}^2N$, the bulk IIB string theory is in the small string
tension and small curvature limit, so that classical string theory is
a good approximation. In this context, the loops are represented in
the bulk by classical (minimal area) worldsheets which end on the
AdS boundary~\cite{Maldacena:Wilson,Rey:Macroscopic}.
For example, two circular Wilson loops of radius $a$, 
which are separated by a distance $L$, correspond in the bulk to a 
worldsheet
with these two loops as its boundary. When the ratio of the size of
the loops to their separation is very small, $a/L\ll 1$, the
worldsheet degenerates into two hemispheres connected by a very thin
tube~\cite{Gross:Aspects}. This degenerate worldsheet represents the 
exchange of light degrees of freedom in the bulk between two
otherwise unaffected minimal surfaces, each with a circle as its
boundary. It is then straightforward, albeit slightly tedious, to
extract the OPE by properly identifying the light states being
exchanged and their coupling to the strings.
For Wilson surfaces, the
strategy is the same, though one now considers membrane worldvolumes 
in M-Theory rather than strings. 
Here we find a new divergent term in the 
calculation which is proportional to the rigid string action. 

The paper is organized as follows. In section~\ref{sec:OPE}, we
examine the OPE for a circular Wilson loop in some detail. We discuss
what operators are allowed to appear in the OPE and list the operators
of low conformal weight explicitly. We compare this with the leading
terms in the perturbative expansion of the Wilson loop. We then
discuss two approaches to the problem of calculating the coefficients
of the OPE at large coupling. The first method consists in 
calculating the correlation function between a Wilson loop and the
various operators of the theory, $\langle W(\CC) \CO_i\rangle$. The
second method is to compute the correlator between two Wilson loops
and to identify the contributions at each order in the
size-to-separation, $a/L$, expansion. 

In section~\ref{sec:circloop}, we find the minimal area string
worldsheet that describes a circular Wilson loop in the 
 fundamental representation. In
section~\ref{sec:lightscalars}, we identify the scalar 
modes that contribute to the supergravity interaction between two
Wilson loops and consider their coupling to the string worldsheets. We
then compute the Wilson loop correlators and extract the 
OPE coefficients. In  addition we 
consider the potential between two rectangular Wilson loops, which is a
straightforward application of the same techniques used to
compute correlators of loops.   
In section~\ref{sec:antisym}, we outline some qualitative features of
the computation of exchange of tensor modes between the worldsheets.

In section~\ref{sec:surface}, we consider 
Wilson surfaces in the $(0,2)$, $d=6$ superconformal theory. We find
the minimal area membrane worldsheet solution describing spherical
Wilson surfaces and the supergravity modes of AdS$_7\times S^4$ which
contribute to their correlation. We find that there is a UV
logarithmically divergent contribution to the area of the surface that
is  proportional to the action of a rigid string embedded
in the 5-brane worldvolume.  

Several details about the bulk-to-bulk and bulk-to-boundary Green's
functions that we will need are presented in an appendix.

\section{The Operator Product Expansion of the Wilson Loop}
\label{sec:OPE}

In~\cite{Maldacena:Wilson,Rey:Macroscopic}, a prescription was given
to compute the effective quark-antiquark potential in the large $N$
strong coupling limit of maximally supersymmetric Yang-Mills
(conformal) field theory in four dimensions. This potential could be
obtained by computing the expectation value of the Wilson loop,
\begin{equation}
\langle W(\CC) \rangle \sim \lim_{\Phi\rightarrow\infty} 
e^{-(S_\Phi-\ell\Phi)}, \label{eq:wilsvev}
\end{equation}
where $\ell$ is the total length of the loop and $\Phi$ is 
a UV regulator~\cite{Susskind:Holo-Bound}.
In this paper, we consider the modified Wilson loop  operator 
given by 
\begin{equation}
W(\CC) = \frac{1}{N} \text{Tr}~P e^{ \oint ds \left[
 i A_\mu(\sigma) \dot{\sigma}^\mu 
+ \theta^I(s)X^I(\sigma)\sqrt{\dot{\sigma}^2}\right]}, 
\label{eq:wilsloop}
\end{equation}
where $A_\mu(\sigma)$ are the four-dimensional gauge fields, the
$X^I(\sigma)$, $I=1,...,6$  are the six scalar fields of 
${\cal N} = 4 $ SYM, and
$\theta^I(s)$ is a point on the five-sphere (so $\theta^2 = 1$). 
As argued in~\cite{Maldacena:Wilson,Rey:Macroscopic} this is the operator
which leads to  a simple calculation in supergravity. The reason is
that, if the gauge symmetry is broken to $U(N-1)\times U(1)$ through a
Higgs VEV, the massive W-bosons can be interpreted as strings
stretched between the horizon and a D3-brane in AdS. These W-bosons
carry, in addition to the charge under the gauge fields, a ``scalar''
charge under the scalar $X^I \theta^I$, where 
$\theta^I$ is the $SO(6)$ orientation of the Higgs VEV, so that we get
the second term in~\eqref{eq:wilsloop}.  

We expect that there exists an operator product
expansion for the Wilson loop when it is probed from  distances large 
compared to its characteristic size $a$,
\begin{equation}
W(\CC) = \langle W(\CC) \rangle \left[ 1 + \sum_{i,n} 
c^{(n)}_i a^{\Delta^{(n)}_i} \CO^{(n)}_i\right], \label{eq:circ-wilsope}  
\end{equation}
where the $\CO^{(n)}_i$ are a set of operators with conformal weights
$\Delta^{(n)}_i$. In this notation, we let $\CO^{(0)}_i$ denote the
$i^{th}$ primary field, while the $\CO^{(n)}_i$ for $n>0$ are its conformal
descendants. For the circular Wilson loop solution, the expectation
value of all operators other than the identity vanishes, so that the
coefficient of the identity is the expectation value of the loop. 

The problem is to explicitly calculate the coefficients $c^{(n)}_i$
that appear in the Wilson loop OPE. In the field theory, this can 
be done perturbatively at weak coupling, but, barring the existence 
of nonrenormalization theorems, the
result cannot, in general, be reliably extrapolated to strong 
coupling. In fact we will see that the coefficients are different.

Local gauge invariant  operators  are  given by traces (over
the gauge group) of polynomials of the scalars $X^I$, the
fermions $\psi^\alpha$, the Yang-Mills field strength $F_{\mu\nu}$,
and their covariant derivatives. Since the operators appearing in the 
Wilson loop must have the same symmetry properties as the Wilson loop
itself, the operators $\CO^{(n)}_i$ should be bosonic and gauge
invariant. 
We consider the case where $\theta^I(s) = \theta^I$ is a constant.  This
breaks the $SO(6)$ R-symmetry of the
superconformal field theory to $SO(5)$, so we  expect that the
$\CO^{(n)}_i$ are $SO(5)$ invariant. Therefore we will obtain
operators which are in irreducible representations of $SO(6)$ that
contain singlets under the $SO(5)$ maximal subgroup. 

We perform the analysis at each conformal dimension. 
\begin{itemize}
\item $\Delta=0$. The only operator of dimension zero is proportional
to the identity.
\item $\Delta=1$. The only elementary fields of dimension one are the
scalars $X^I$. A gauge invariant operator would be $\Tr(X^I)$, but 
taking $SU(N)$ as 
the gauge group~\cite{Witten:Holography}, this trace 
vanishes.
\item $\Delta=2$.
At dimension two, the only trace over the gauge group that is
non-trivial is that of the scalar bilinear $\Tr(X^I X^J)$, which
splits in two irreducible representations of $SO(6)$. One is the
singlet $\Tr(X^I)^2$ which
 is expected to have a large anomalous dimension~\cite{Witten:Holography},
 so it
must be dropped from the expansion at low orders in the supergravity regime.
The other is the symmetric traceless
tensor, $\CC^A_{IJ}\Tr(X^I X^J)$, which is the $\mathbf{20}$ of
$SO(6)$. Its conformal dimension is independent of the
 coupling~\cite{Witten:Holography}. Since the
$\bf20\rightarrow 1\oplus 5\oplus 14$ under $SO(5)$, we find that the
operator in the $SO(5)$ singlet, {\it i.e.}, the projection of
$\CC^A_{IJ}\Tr(X^I X^J)$ will appear in the
OPE.  
\item $\Delta=3$.
At dimension three, we must consider the scalar operator 
$\Tr (X^I X^J X^K)$.  Once again, only the symmetric
traceless part, $\CC^A_{IJK}\Tr (X^I X^J X^K)$
(in the $\mathbf{50}$), has a protected conformal dimension.
 All other components ought  to have large anomalous 
scaling  dimensions.

We must also consider the scalar bilinear in the fermions, 
$\Tr (\psi^\alpha_A \psi_{B\alpha})$, where $A$ and $B$ are spinor
indices for the R-symmetry group. This operator
is in the $\mathbf{10}\rightarrow\mathbf{10}$ and does not contain 
any $SO(5)$ singlets, so it will not contribute. 

We can also have $\Tr (X^I F_{\mu\nu} )$, which transforms as a two 
form under Lorentz transformations and decomposes as the 
$\mathbf{6}\rightarrow \mathbf{1}\oplus\mathbf{5}$ under $SO(5)$ and
should contribute to the OPE. 
For a circular Wilson loop, 
the allowed components depend on the orientation
$\hat{\sigma}^{\mu\nu}$ of the loop. 

The final primaries at this order are the R-symmetry currents,
$J^{[IJ]}_{\mu}$, which are in the adjoint of $SO(6)$, which is the
antisymmetric tensor, $\mathbf{15}$. Under $SO(5)$,
$\mathbf{15}\rightarrow \mathbf{5}\oplus\mathbf{10}$, so there is no
$SO(5)$-invariant component. Therefore this operator does not
contribute to the OPE.

Finally, we can also have $\CC^A_{IJ} \partial_i \, \Tr(X^I X^J)$, which 
is a superconformal descendant of $\CC^A_{IJ} \Tr(X^I X^J)$. 
In the particular case of a circular Wilson loop OPE, it is forbidden
by rotational invariance. 

\item $\Delta=4$.
At dimension four, there are various chiral primaries. The operators
which contain 
$SO(5)$ singlets and should appear in the OPE are the symmetric
traceless rank 4 tensor $\CC^A_{IJKL}\Tr (X^I X^J X^K X^L)$, the field 
strength
operator $\CC^A_{IJ}\Tr (X^I X^J F_{\mu\nu} )$ ,
the energy-momentum tensor, and 
the Lagrangian. Additionally, one 
can have descendents of the $\Delta=2,3$ operators which
 already appeared above, as well as two trace operators like 
$\CC^A_{IJ}\Tr (X^I X^J)\, \CC^B_{KL}\Tr( X^K X^L)$.
In the 't Hooft limit, we expect to find only single trace operators. 
\end{itemize}

Summing up our results, we arrive at the following expression for the
circular Wilson loop OPE
\begin{equation}
\begin{split}
\frac{ W(\CC) }{\langle W(\CC) \rangle }
 &=   \Bid 
+ c^{(0)}_2 \, a^2 \, Y^{(2)}_A(\theta) {\cal N}_2  \CC^A_{IJ} 
\Tr\left(X^I X^J\right) 
+ c^{(0)}_3 \, a^3 \, Y^{(3)}_A(\theta) {\cal N}_3 \CC^A_{IJK} 
\Tr\left(X^I X^J X^K\right)  \\
& \hspace*{1.5cm}  + c^{(0)}_4 \, a^3  \hat{\sigma}^{\mu\nu} 
\Tr\bigl(\theta^I X^I  F_{\mu\nu} \bigr)   + \cdots ~,
\end{split} \label{eq:low-ord-exp-circ}
\end{equation}
where $\hat\sigma^{\mu\nu}$ is a unit two-form which denotes the
orientation of the Wilson loop,  ${\cal N}_k$ is a constant that 
ensures that operators are ``unit'' normalized, in a sense indicated
below and $C^A_{I_1\cdots I_k}$ is a basis of symmetric traceless
tensors such that the spherical harmonics are 
$Y^A(\theta) = C^A_{I_1\cdots I_k}\theta^{I_1}\cdots\theta^{I_k}$, 
with the index $A$ running over all the spherical harmonics
of given $SO(6)$ Casimir, see~\cite{Lee:3point} for conventions. 
 
\subsection{Perturbative Calculation of the OPE Coefficients}

For small $\lambda=g_{YM}^2 N$, we can perturbatively
expand the Wilson loop~\eqref{eq:wilsloop} to find an expression for the
OPE~\eqref{eq:low-ord-exp-circ}. We find the first few terms to be
\begin{equation}
\begin{split}
\frac{ W_{\text{pert}}(\CC) }{\langle W_{\text{pert}}(\CC) \rangle }
 = & 1 
+ \sum_{k\geq 2} \frac{(2\pi  a)^k}{k!\, N} 
\theta^{I_1}\cdots \theta^{I_k}
\Tr\left(X^{I_1} \cdots X^{I_k} \right)  \\
& - \frac{2\pi^2 a^3}{N} \theta^I \hat{\sigma}^{\mu\nu} 
\Tr\bigl(X^I  F_{\mu\nu} \bigr)
-\frac{\pi^2a^4}{2N} \Tr(\hat{\sigma}^{\mu\nu}F_{\mu\nu})^2 + \cdots.
\end{split} \label{eq:pert-ope}
\end{equation}
Some of these operators will get high conformal dimensions in the 
strong coupling limit and so we know that they will not appear
as leading terms in the expansion. There are, however, operators
whose dimensions are protected, such as the symmetric
traceless combinations  $C^A_{I_1...I_k}\Tr(X^{I_1}... X^{I_k})$. 
The operator product coefficients will depend on the normalization
of the operators. We will choose them to be ``unit'' normalized, in the
sense that 
$\langle \CO(\vec{x})\CO(\vec{y}) \rangle
= |\vec{x}-\vec{y}|^{-2\Delta}$.  The operators
in~\eqref{eq:pert-ope} are not normalized. In order to normalize them,
we have to compute their two-point functions. These were calculated 
in~\cite{Lee:3point} and using those results, we find that 
the operator product coefficients for the highest weight chiral primaries
have the behavior\begin{equation}
c_\Delta \sim  \frac{\lambda^{\Delta/2}}{N}. \label{eq:pert-coeff}
\end{equation} 
where  $\lambda=g_{YM}^2N$. For the other protected  operators,
we find a similar $\lambda^p/N$
dependence, where the exponent $p$ is related to the conformal weight
of the operator. We will see that for large $\lambda$, the dependence
on $N$ will be the same, but that the $\lambda$ dependence is different. 
Of course the dependence on $N$ can be understood in a simple fashion
by using large $N$ counting arguments.

\subsection{Supergravity Calculation of the OPE Coefficients}

We now describe how to  calculate the coefficients in the 
supergravity description. 
There are two ways to determine the coefficients of the operator
product expansion~\eqref{eq:low-ord-exp-circ}. The first and most
straightforward method 
is to compute the correlator of the Wilson loop with each operator
that is expected to appear in the loop. This correlator gets
contributions only from the given conformal primary and its
descendents, 
\begin{equation}
\frac{ \langle W(\CC) \CO^{(0)}_i \rangle }{ \langle W(\CC) \rangle }
= c^{(0)}_i \frac{a^{\Delta^{(0)}_i}}{L^{2\Delta^{(0)}_i}} 
+\sum_{m > 0} c^{(m)}_i a^{\Delta^{(m)}_i}
\langle \CO^{(m)}_i \CO^{(0)}_i \rangle. \label{eq:corr-singop}
\end{equation}
Here we have isolated the contribution from the descendents and their
mixings with the primaries in the second term.

\begin{figure}[h]
\centerline{\epsfxsize=6cm \epsfbox{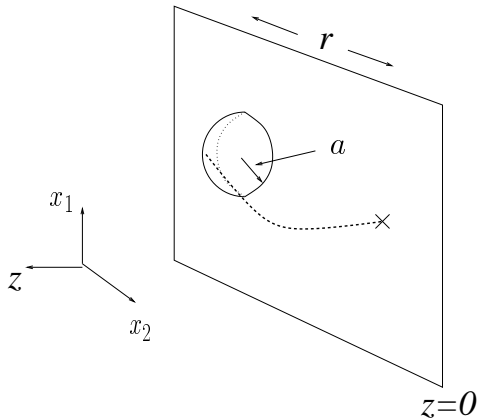}}
\caption{The emission of a particle from a circular Wilson loop of
size $a$ onto the AdS boundary at a distance $r$ from the loop.}
\label{fig:circwilsop}
\end{figure}

In the supergravity description, the Wilson loop will be related to 
a string worldsheet ending on the boundary of AdS.
The correlation function~\eqref{eq:corr-singop} can be calculated by
treating the string as an external source for the fields propagating
in anti-de Sitter spacetime and then computing the string effective
action for the emission of supergravity states onto the point on the
boundary where the operator is inserted. See  Figure~\ref{fig:circwilsop}.

Another approach to the problem of calculating the OPE coefficients 
is to compute the correlator of a pair of Wilson loops that are separated 
by a distance which is large compared to their size. In the conformal
field theory, the correlator can be calculated from the operator
product expansion for the two Wilson loops
\begin{equation}
\begin{split}
\frac{ \langle W(\CC,L) W(\CC,0) \rangle }{ 
\langle W(\CC,L) \rangle \langle W(\CC,0) \rangle  }  = &
\sum_{i,j;m,n} c^{(m)}_i c^{(n)}_j \, 
a^{\Delta^{(m)}_i+\Delta^{(n)}_j}
\langle\CO^{(m)}_i(L)\CO^{(n)}_j(0)\rangle \\
= &  \sum_i \bigl(c^{(0)}_i\bigr)^2 \, 
\frac{a^{2\Delta^{(0)}_i}}{{L^{2\Delta^{(0)}_i}}} \\ 
& + \sum_{i,\{m,n\}\neq\{0,0\}} c^{(m)}_i c^{(n)}_i \, 
a^{\Delta^{(m)}_i+\Delta^{(n)}_i}
\langle\CO^{(m)}_i(L)\CO^{(n)}_i(0)\rangle.
\end{split} \label{eq:corr-twoloops}
\end{equation} 
In the last line, the first term is due solely to the primary fields,
while the second contains the contributions from descendents.

\begin{figure}[h]
\centerline{\epsfxsize=6cm \epsfbox{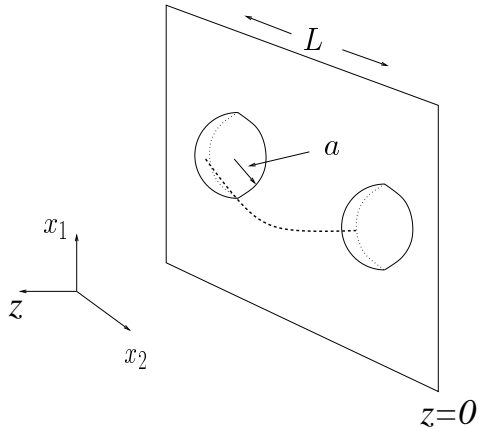}}
\caption{The correlation between two circular Wilson loops.} 
\label{fig:circwilsint} 
\end{figure}

In  the supergravity approximation, these Wilson loop correlators can
be calculated by computing the amplitude for the exchange of
light states between the two string worldsheets which have the Wilson
loops as their boundaries~\cite{Gross:Aspects}, as represented in
Figure~\ref{fig:circwilsint}. We will actually calculate the OPE coefficients
in this fashion, since it is slightly simpler. 

In the next section, we will examine the details of the circular
Wilson loop solution. With the necessary information in hand, we will
then return to the computation of correlation functions of Wilson
loops. 

\section{Circular Wilson Loops and AdS Supergravity}
\label{sec:circloop}

According to~\cite{Maldacena:Wilson}, in order to compute
the expectation value of a circular Wilson loop in the 
large $g_s N$ limit, we should find the minimal area 
string worldsheet ending on a circle at the boundary of anti-de Sitter
space. We choose the scalar charge of the Wilson loop to be constant,
so that the string worldsheet lives at a single point on the
5-sphere. This implies that $\theta^I(s) = \theta^I$
in~\eqref{eq:wilsloop} is a constant. We could find the classical 
worldsheet by solving the Euler-Lagrange equations coming from the
Nambu-Goto action in this background, however, in this case there is 
a simpler way to find the worldsheet.

We note that the Euclidean conformal group in 4 dimensions, 
$SO(1,5)$, has
elements which map straight lines into circles, namely the special
conformal transformations,
\begin{equation}
x^{\prime i} = \frac{ x^i + c^i x^2}{1+2 c\cdot x + c^2 x^2},
\label{eq:spec-ct} 
\end{equation} 
where $c^i$ is a vector in $\BR^4$. We take the AdS metric
\begin{equation}
ds^2 =  \frac{1}{z^2}\left( dz^2 + dx^i dx^i \right),
\end{equation}
with the boundary at $z =0$. The special conformal
transformation~\eqref{eq:spec-ct} corresponds to the AdS 
reparameterization 
\begin{equation}
\begin{split}
x^{\prime i} &= 
\frac{ x^i + c^i \left( x^2 + z^2  \right) }{
1 + 2 c\cdot x + c^2 ( x^2 + z^2 )} \\
z^\prime &= \frac{z}{ 1+2 c\cdot x + c^2 \left(x^2+ z^2\right) } .
\end{split} \label{eq:AdS-ct}
\end{equation}

Starting with a line on the boundary and a minimal area
surface in AdS which ends on 
that line (which just extends along the line and the $z$ direction),
 we can apply the above conformal transformation and map it to
a circle in such a way that on the boundary we have a circle of radius
$a$. Then the surface in AdS is given by\footnote{While
 this paper was being written, we learned that
D.\ Gross has independently obtained this circular Wilson loop
solution~\cite{Gross:Strings-98}.}. 
\begin{equation}
x^{\prime i} = \sqrt{a^2 - z^2 } 
\left( e_1^i\cos\psi + e_2^i \sin\psi \right),~~~  
0 \leq z \leq  a,~0\leq\psi < 2\pi. 
\end{equation}
where $e_1, e_2$ are two orthonormal vectors on the boundary. 
We see that this surface ends on the boundary on a circle of radius $a$ and
it closes off at $ z = a$. 
It is  useful  to compute the area element,
\begin{equation}
d{\cal A } =  dz \, d\psi \sqrt{\det\left(g_{\mu\nu}(x)
\partial_\alpha x^\mu \partial_\beta x^\nu\right)}
=  {a \, dz \, d\psi \over z^2 } . \label{eq:darea}
\end{equation}
As in~\cite{Maldacena:Wilson}, we find that there is a divergence in
the action, 
\begin{equation}
S_\epsilon  = 
\frac{1}{2\pi\alpha^\prime} \int_\epsilon^{a} dz \int_0^{2\pi} 
d\psi \, \frac{ az }{z^3} 
= \frac{1}{\alpha'}\left(  \frac{a}{\epsilon} - 1 \right).  
\label{eq:loopact}
\end{equation}

In terms of the theory on the boundary, this divergence corresponds
to the UV divergence in the Coulombic self-energy of a point
charge. In the bulk theory, this  divergence is  due to the
contribution of an infinitely long straight string ending on the
circle. 
After subtracting the divergent term, we find that $S=-1/\alpha'$, which 
is independent of the radius $a$ of the loop, as required
by conformal invariance.
We are choosing units in which the radius of AdS is equal to one, so that
$\alpha' = 1/\sqrt{ 4 \pi g_s N}$. 

\section{Contributions from the Lightest Scalars}
\label{sec:lightscalars}

We will be primarily interested in the contributions from the 
lightest scalars, whose exchange will dominate the long distance
interactions. These light states correspond to the 
operators of lowest dimension in the OPE for the Wilson loop.
The relevant modes may be determined from the KK mass  
spectrum listed in Figure~2 or Table~III of~\cite{Kim:KK-spectrum}. 

\subsection{The Dilaton}

We will start calculating the coefficient of the OPE in front of the 
operator associated to the dilaton. We start with this field because
the calculation of the precise coefficient is simpler. We will consider
other cases later. The dilaton can be  expanded in Kaluza-Klein
harmonics as 
\begin{equation}
\phi=\sum_k \phi^k Y^{(k)}(\theta).
\end{equation}
The action for the dilaton is  
\begin{equation}
\begin{split}
S_{\text{kin.}} &= 
\frac{1}{2\kappa_{10}^2} \int d^{10}x \, 4 (\nabla  \phi)^2 \\
&=  \int d^{5}x\, \sum_k B^\prime_k \, 
\frac{1}{2}\left[ (\nabla\phi^k)^2
+ m_k^2 {\phi^k}^2 \right] ,
\end{split}
\end{equation}
where, in units where $R_{\text{AdS}}=1$,  
\begin{equation}
B^\prime_k = { N^2 \over 2^{k-1} \pi^2 (k+1)(k+2) },
\end{equation}
and  we are normalizing the spherical harmonics as
in~\cite{Lee:3point}.  

The coupling of the dilaton to the string worldsheet can be found
remembering that the supergravity calculations were done using the
Einstein metric, while the string worldsheet couples to the
string metric, $G_{string} = e^{\phi/2} g_{Einstein}$. So the coupling
of the string worldsheet to the dilaton is given by 
\begin{equation}
S_{\text{dil}}= \frac{1}{2\pi\alpha^\prime} \int d^2\sigma \sqrt{\gamma}\,
e^{\frac{\phi}{2}} 
+\frac{1}{4\pi}\int d^2\sigma \sqrt{\gamma}R^{(2)} \phi ,
\end{equation}
where $\gamma_{\alpha\beta}$ is the induced metric on the 
worldsheet and $R^{(2)}$ 
is the corresponding worldsheet curvature. 
In the case that the radius of $AdS$ is large, the term involving the
worldsheet curvature will be subleading compared to the first term.
Therefore we should neglect the curvature contribution in the
amplitudes we consider.  

So now we want to calculate the contribution of the dilaton to the 
expectation value of the product of Wilson loops.
This contribution will be given by
\begin{equation}
{ \langle W W \rangle_{dil}
\over \langle W \rangle^2 } = {\rm Exp}
\left[ \sum_{k, A} Y^{(k)}_{A}(\theta)^2 \frac{1}{4}
\int \frac{d{\cal A}}{2 \pi \alpha'}
 \int \frac{d {\cal A}'}{2 \pi \alpha'} 
G_k(W(\sigma,\sigma'))_{dil} \right],
\end{equation}
where the sum over $A$ indicates the sum over all spherical harmonics
of total angular momentum $J^2 = k(k+4)$~\footnote{
This sum gives a result that is
independent of $\theta$, but we leave the result in this form to read off
the contribution to the OPE of each Kaluza-Klein mode.}.  
This formula summarizes the effects of dilaton exchanges between
the two worldsheets. Of course, the dominant term for large $N$ is
 the one dilaton exchange arising from the first-order expansion of the
exponential. Notice that  $\sigma,\sigma'$ indicate points on
the two separated worldsheets. If the separation between the worldsheets
is very large, then $W$ is very small and we can approximate the Green's 
function~\eqref{eq:UHS-Greens} by
\begin{equation}
G_k(W)_{dil}  \sim  { \alpha_0 \over {B'_k}} W^{\Delta} 
\sim  { \alpha_0 \over {B'_k}}
{ z^\Delta {z'}^{\Delta} \over L^{2\Delta} },
\label{eq:approxgreen}
\end{equation}
where $L$ is the (large) distance between the two loops,
 $ \Delta = 4 +k$ and $\alpha_0 = (\Delta -1)/2\pi^2$.
Using~\eqref{eq:darea} and~\eqref{eq:approxgreen} we find that the integrals
over the worldsheet reduce to simple 
expressions involving $\int dz \, z^{\Delta -2}$
which are always convergent at $z=0$. The final result is that 
\begin{equation}
{ \langle W W \rangle_{dil} \over \langle W \rangle^2} 
\sim \sum_{k, A} \frac{2^{k-2} \pi (k+1)(k+2)}{k+3} \, 
\frac{g_s}{N} \, Y^{k}_{A}(\theta)^2 \left(\frac{a}{L}\right)^{2\Delta},
\label{eq:opedil}
\end{equation}
so that
\begin{equation}
c_{dil,\Delta } 
= 2^{\Delta/2-3}  \sqrt{\frac{\pi(\Delta-2)(\Delta-3)}{\Delta-1}} 
\frac{ \sqrt{g_sN}}{N}  
\end{equation}
Notice that we are unit-normalizing the dilaton operator.

Of course we could  have calculated this last result directly by 
computing the one point function of the operator associated to
the dilaton in the presence of a Wilson loop. 
In fact, as shown in~\cite{Vijay:probes}, the one point function is
proportional to the value of the dilaton  near the boundary. This
value is non-zero because the string worldsheet acts a source for the
dilaton. We can actually calculate the correlation function of the
operator with the Wilson loop for any position, not just for large
distances. We get
\begin{equation} 
\begin{split}
\frac{ \langle W \CO^k(\vec x) \rangle}{ \langle W \rangle } & =  
C_{\Delta} \lim_{z \to 0} z^{ -\Delta} \phi^k(z,\vec x) 
= (const.) Y^{(k)}(\theta) { 1 \over 2 \pi \alpha' } \int d{\cal A}' 
{{z'}^{\Delta} \over [(\vec x - \vec {x'})^2 + {z'}^2 ]^{\Delta} } 
\\ 
& = c_{dil,\Delta } Y^{(k)}(\theta) 
\frac{a^{2 \Delta} }{\left[ ( y^2 + r^2 - a^2)^2 + 4 a^2 y^2 
\right]^{\Delta/2} },
\label{eq:full} \end{split} \end{equation}
where $r$ is the polar coordinate on the plane defined by the loop and
$y$ is the distance on the plane orthogonal to the loop. 
We see that when the operator approaches the loop
we have a singularity of the form $ 1/d^{\Delta}$ where $d$ is the distance
from the loop. In fact, the result~\eqref{eq:full} could be derived in a 
simple way by first computing $\langle W \CO \rangle $ for a straight line 
and then applying the conformal transformation mapping the line to a
circle.  This result~\eqref{eq:full} includes all the information about the
conformal descendents of the operator $\CO$ appearing in the OPE.  

\subsection{``Tachyonic'' Scalars}

The leading term in the OPE expansion of the Wilson loop comes
from an operator of dimension two. This operator comes from 
a field with $m^2 <0 $ in the supergravity theory. 
Scalars arise from several supergravity fields. 
There are  scalar
KK modes of the metric over the 5-sphere, (in the notation
of~\cite{Kim:KK-spectrum}) 
\begin{equation} 
h^\alpha_\alpha(x,\theta) = \sum_k \pi^k(x) Y^{(k)}(\theta), 
\end{equation}
as well as the scalar KK modes of the antisymmetric 4-form, 
\begin{equation}
a_{\alpha\beta\gamma\delta}=\sum_k b^k
{\epsilon_{\alpha\beta\gamma\delta}}^\epsilon D_\epsilon Y^{(k)}(\theta).
\end{equation}  
One linear combination of these has tachyons in its spectrum.
Another scalar comes from the trace of the metric on the AdS
component, $h^{ \mu}_\mu$.
 We can algebraically express these in terms of the
$\pi^k$, as  
\begin{equation}
\begin{split}
 \sum_k H_{\mu\mu}^k Y^{(k)}(\theta) &\equiv h_\mu^{\mu} +
 { 5 \over 3} h_{\alpha}^\alpha \\ 
&= \frac{16}{15} \sum_k \pi^k Y^{(k)}(\theta),
\end{split} \label{eq:h-in-h}
\end{equation}
so they contribute to the states of this family. 

From the field equations, given as equation~(2.33)
of~\cite{Kim:KK-spectrum}, one sees that the modes $\pi^k$ and $b^k$
mix. The mixing 
angles, as well as the normalized action for the mass eigenstates has
been conveniently presented by Lee~{\it et.\ al.}~\cite{Lee:3point}. 
They found that the mass eigenstates were 
\begin{equation}
\begin{split}
& s^k = \frac{1}{20(k+2)} \left[ \pi^k - 10(k+4) b^k \right], 
\hspace*{2mm} M_s^2 = k(k-4) , 
\hspace*{11.5mm} k\geq 2, \\
& t^k = \frac{1}{20(k+2)} \left[ \pi^k + 10k b^k \right], 
\hspace*{12mm}M_t^2 = (k+4)(k+8),
\hspace*{2mm}  k \geq 0.
\end{split} \label{eq:eigenstates}
\end{equation}
The lightest states correspond to the lowest modes of $s^k$, so we
will focus on these modes. The action for the $s^k$ was found to
be~\cite{Lee:3point} 
\begin{equation}
S = \int d^5x \sum_k B_k \,  \frac{1}{2} 
\left[ (\nabla_\mu s^k)^2 + k(k-4) (s^k)^2 \right],
\end{equation} 
where  
\begin{equation}
B_k = { 2^{3-k} N^2 k(k-1) \over \pi^2 (k+1)^2 },
\end{equation}
and our normalizations are as in~\cite{Lee:3point}.

In order to compute the coupling of  $s$
 to the string worldsheet, we need to
find all the supergravity fields that are excited when $s$ is nonzero and
all the other ``diagonal'' modes are set to zero. 
From the equations that we saw above we find a contribution to 
$h_\mu^\mu$ and there is also a contribution to 
\begin{equation}
H_{(\mu \nu)} = { 4 \over k+1 } D_{(\mu}D_{\nu)} s ,
\label{eq:bigh} \end{equation}
where the parenthesis indicate the symmetric traceless combination.

Now we should find how $s$ couples to a string worldsheet. 
These couplings will involve terms with derivatives. 
In the calculation we are interested in these derivatives 
will act on the Green's function of the field $s$. 
Moreover, we will be interested in extracting the leading piece in 
$L$, where $L$ is the separation between two Wilson loops or a Wilson
loop and an operator. In that regime we will be able to approximate the
Green's function by an expression like $ z^{\Delta}/[
 ( \vec x -\vec x')^2 +z^2]^\Delta $, so that the only derivatives which
will not produce a subleading term in $1/L$ will be those acting on the
numerator of this expression. This implies that in calculating
the coupling to $s$ from~\eqref{eq:bigh} we will be able to replace
$z$-derivatives by factors of $\Delta$, etc. 
The string worldsheet will couple to the various components of the metric
on $AdS_5$. Couplings through $a_{\alpha\beta\gamma\delta}$ will be
small if $R$ is large, since they will involve couplings to the
worldsheet fermions. 
Finally we get the coupling to the worldsheet as 
\begin{equation}
{ 1 \over 2 \pi \alpha'} \int d{\cal A} { (-2 k s_k ) }
{ z^2 \over a^2 }. \label{eq:s-coupling} 
\end{equation}
By using the same method as we used for the dilaton we obtain
\begin{equation}
\frac{ \langle W W \rangle_{s} }{  \langle W \rangle^2  }  \sim  
\sum_{k,A} {  \alpha_0 \over B_k } \left( {2 k \over 
(k+1) \alpha' } \right)^2 
Y_{A}^{(k)}(\theta)^2 { a^{2 k } \over L^{2 k} }   
= \sum_{k,A} { 2^k  \pi k } \, {g_s  \over N } \, 
Y^{(k)}_{A}(\theta)^2 { a^{2 k } \over L^{2 k} }.   
\end{equation}
Similarly, we can calculate 
\begin{equation}
\frac{\langle W \CO^k_A \rangle  }{  \langle W \rangle }
\sim 2^{k/2} \sqrt{ \pi k} \,\sqrt{{g_s \over N} } 
Y^{(k)}_A(\theta) { a^{k} \over L^{2 k} }.
\end{equation}
From these expressions we determine the OPE coefficients
\begin{equation}
c_{s,\Delta} = 2^{\Delta/2-1} \sqrt{ \Delta} \, \frac{\sqrt{\lambda}}{N}.
\end{equation}
This equation should be compared to the weak coupling
 result~\eqref{eq:pert-coeff}. As expected
 the $N$ dependence is the same but the powers of 
 $g^2_{YM} N$  are different. This is no contradiction, since the
two calculation have different regimes of validity.

\subsection{The Potential Between Two Rectangular Wilson Loops}

\begin{figure}[h]
\centerline{\epsfxsize=7cm \epsfbox{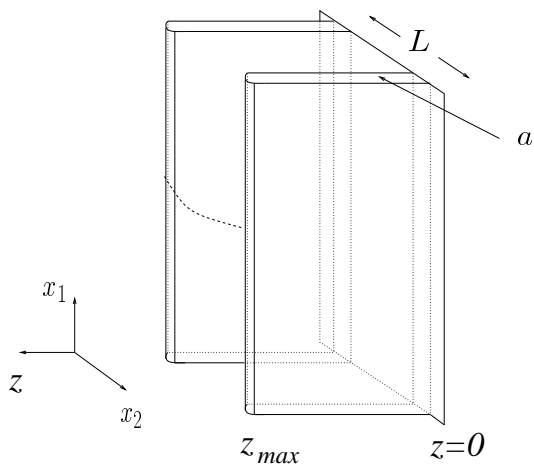}}
\caption{The interaction between two rectangular Wilson loops of size
$a$, separated by a distance $L\gg a$.} 
\label{fig:wilsint} 
\end{figure}

The tools that we have collected to compute correlation functions due
to exchange of scalar supergravity modes between string worldsheets
are also applicable to the study of the potential between rectangular
Wilson loops.
For a pair of rectangular
Wilson loops, each of size $a$,
which are separated by a distance $L\gg a$, as depicted in
Figure~\ref{fig:wilsint}, the potential takes the
form~\cite{Peskin:Diagrammatics}  
\begin{equation}
\begin{split}
V=& - \lim_{T\rightarrow\infty} 
\frac {\log \langle W(\CC,L) W(\CC,0) \rangle 
-  \log\langle W(\CC,L)\rangle \langle W(\CC,0)\rangle }{T}  \\
=&  \sum_n V_n \frac{a^{2n-2}}{L^{2n-1}} .
\end{split} \label{eq:exp-prod} 
\end{equation}
Perturbatively, we will find that the $V_n\sim \lambda^p/N^2$. At
large $\lambda$, the $N$ dependence is the same, but the $\lambda$
dependence will be different. 

We will use the worldsheet solutions
of~\cite{Maldacena:Wilson} to compute the effective
potential. We find that,  for large separations, the
asymptotic behavior of the scalar Green's
function~\eqref{eq:UHS-Greens} is 
\begin{equation}
G_k(x,x^\prime) = { \alpha_{0,k} \over A} 
\left(  
\frac{ z z'}{L^2 + (\tau-\tau^\prime)^2} \right)^\Delta + \cdots.
\label{eq:green-approx}
\end{equation}
The surface  was given by 
\begin{equation}
x(z) = z_{max} \int_{z/z_{max}}^1 dy \frac{{y}^2}{\sqrt{1 - y^4 }},
\end{equation}
where $z_{max}$ is determined by the condition that 
$|x(0)-x(z_{max})| = a/2$, so that
$z_{max} = a \Gamma(1/4)^2/( 2 \pi)^{3/2} $. 
The area element is given by 
\begin{equation}
d{\cal A} =  dt \, dz \, 
\frac{ z_{max}^2  }{z^2 \sqrt{ z_{max}^4 - z^4 } }.
\end{equation}
The coupling to the field $s$ is given by 
\begin{equation}
S = {1 \over 2 \pi \alpha'} \int d{\cal A} 
{(-2 k s) z^2 \over z_{max}^2 },
\end{equation}
where we used the same method as above to calculate the 
coupling involving terms with derivatives of $s$. 
So the final expression for the potential between two Wilson loops is
\begin{equation}
\begin{split}
V^{({s})} =& -\frac{S^{({s})}}{T} \\
& = - \sum_{k,A} 
{ \alpha_0 \over B_k } \,
{ \Gamma(k-1/2)\sqrt{\pi} \over \Gamma(k) L^{2k-1} } \,
\left[ \frac{( -2 k) z_{max}^{k-1}}{2 \pi \alpha'}\,
{ \sqrt{\pi} \Gamma( \tfrac{k+1}{4}) \over 
  4 \Gamma(\tfrac{k+3}{4}) } \right]^2 Y^{(k)}_{A}(\theta)^2  \\ 
& = - \sum_{k} 
\frac{ \Gamma(\tfrac{1}{4})^{4k-4}}{3 2^{2 k+5}   \pi^{3k-7/2} } \,
\frac{k(k+1)^2(k+2)(k+3)
\Gamma(k-\tfrac{1}{2})\Gamma(\tfrac{k+1}{4})^4}{
\Gamma(k)\Gamma(\tfrac{k+1}{2})^2} \,
\frac{g_sN}{N^2}  \frac{a^{2k-2}}{L^{2k-1}}.
\end{split} \label{eq:eff-pot}
\end{equation}
We have done the sum $\sum_A Y^k_A(\theta)^2 =  (k+3)(k+2)/ (3\cdot 2^{k+1})$.
We    have included contributions to the potential 
coming from all of the Kaluza-Klein modes of the field $s$. The ones
with $k=2$ will give the leading contribution. Of course other fields
will also make contributions to the potential which are comparable to
the terms above with $k\geq 3 $. For example, we can calculate
the contributions to the potential from the dilaton 
\begin{equation}
\begin{split}
V^{\phi} =&  - \sum_{k,m_k} 
{ \alpha_0 \over {B'_k} } \,
{ \Gamma(\Delta -1/2)\sqrt{\pi} \over \Gamma(\Delta) 
 L^{2\Delta-1} } \,
\left[ \frac{ ( 1/2) z_{max}^{\Delta-1}}{2 \pi \alpha'} 
{ \sqrt{\pi} \Gamma( \tfrac{\Delta-1}{4}) 
\over    4 \Gamma(\tfrac{\Delta+ 1}{4}) } \right]^2 
Y^{(k)}_{m_k}(\theta)^2  \\ 
=& -\sum_{k} \frac{ \Gamma(\tfrac{1}{4})^{4\Delta-4}}{ 
3 2^{2 \Delta+9}  \pi^{3\Delta-7/2} } \,
\frac{(\Delta-1)^2(\Delta-2)^2(\Delta-3)
\Gamma(\Delta-\tfrac{1}{2})
\Gamma(\tfrac{\Delta-1}{4})^4}{
\Gamma(\Delta)\Gamma(\tfrac{\Delta-1}{2})^2} \\
& \hspace*{2.5cm} \cdot \frac{g_sN}{N^2}  
\frac{a^{2\Delta-2}}{L^{2\Delta-1}},
\end{split} 
\end{equation}
where now $\Delta = 4 +k $. As $g_s\sim \lambda/N$,
these results agree with the behavior illustrated
in~\eqref{eq:exp-prod}.  

\subsection{Contributions from Vectors and Tensors}
\label{sec:antisym}

Let us finally note that the operator product coefficients
involving other operators can be computed in a similar way. They will
involve the contribution of various supergravity fields to the
correlator between two different Wilson loops. In particular, the term
$\text{tr}(X^I F_{\mu\nu})$ term in the
OPE~\eqref{eq:low-ord-exp-circ} corresponds to the lowest mode of an
antisymmetric tensor, $B_{\mu\nu}$, on AdS$_5$. 

\section{The Spherical Wilson Surface and AdS$_7$}
\label{sec:surface}

In~\cite{Maldacena:Wilson}, it was shown that one could use the
AdS description of the large~$N$ limit of the $(0,2)$ superconformal
field theory in six dimensions to compute Wilson surface
observables~\cite{Ganor:largeN}, even though an explicit formulation
of the field theory does not exist, so that there is no formula
analogous to~\eqref{eq:wilsloop}. 
Let us consider a spherical Wilson surface. We take the scalar charge
of the surface to be constant (a point on $S^4$). 
In the gravity picture, the Wilson surface should be the 
boundary of a  minimal area membrane worldvolume in AdS$_7\times S^4$.
One can either  solve the equations of
motion directly to
obtain the minimal worldvolume, or, by analogy with the discussion of 
section~\ref{sec:circloop}, one can note that a flat plane in the
boundary of AdS$_7$ can be conformally mapped to a sphere. This flat 
plane is the boundary of an infinite membrane that is stretched
between the AdS boundary and $z = \infty $. The conformal mapping maps
the worldvolume into a 3-hemisphere whose boundary is a 2-sphere that
corresponds to the CFT Wilson surface.
A convenient parameterization of the solution is given in terms of 
the
Poincare  coordinates as
\begin{equation}
\begin{split} 
x_1&=\sqrt{a^2 - z^2} \cos\theta\\
x_2&=\sqrt{a^2 - z^2} \sin\theta\cos\psi\\
x_3&=\sqrt{a^2 - z^2} \sin\theta\sin\psi,
\end{split} \label{eq:surf-param}
\end{equation}
where $0\leq z \leq a$, $0\leq\theta\leq \pi$, and $0\leq \psi\leq 
2\pi$. Now we take the radius of $AdS$ to be equal to one, then 
the radius of $R_{S_4} =1/2$, $l_p = 1/(8 \pi N)^{1/3}$, and the
tension of the two-brane is $T^{(2)} = 1/(2 \pi)^2 l_p^3 = 2 N/\pi $.

We then find that the volume  of the membrane is divergent
\begin{equation}
\begin{split}
S &= T^{(2)} \int d {\cal V} =    
T^{(2)}  4 \pi \int_\epsilon^a \frac{ dz \, a \sqrt{a^2 - z^2 } }{ z^3}\\
& =  \pi T^{(2)}  \left[ 
+ \frac{2 a^2}{\epsilon^2 } - 2 \ln\frac{2 a }{\epsilon} -1 + 
{\cal O}(\epsilon)
 \right] .
\label{eq:div-mem-act} 
\end{split}
\end{equation}
We can make several observations  regarding   this expression for the
action. First, we see that the action scales as $N$, 
in agreement with the scaling found for the ``rectangular'' solution
of~\cite{Maldacena:Wilson}.
As indicated in~\cite{Susskind:Holo-Bound}, $\epsilon$ should be thought of
as a UV cutoff. So we see that we have two divergent terms. The quadratic
divergence is proportional to the area of the surface. This term
was also present in the case of a rectangular Wilson 
surface~\cite{Maldacena:Wilson}. In this case, we see that there is also 
a logarithmic divergence. The first question is what would this divergence
be in a more general case? It can be seen, by analyzing the equations
of motion of the theory, that 
for a generic two-dimensional surface $\Sigma_2$ 
this logarithmic divergence is proportional to the ``rigid string'' action
\cite{polyakov} 
\begin{equation}
S_{rigid} = \int_{\Sigma_2}  d^2 \sigma  \sqrt{\gamma} ( \nabla^2 X^i)^2,
\label{eq:rigid} 
\end{equation}
where $\gamma$ is the induced metric on the Wilson surface and $X^i$ are
the coordinates on $\BR^6$ describing the surface. Notice that $\Sigma_2$
is a  surface in  the boundary six-dimensional 
field theory. As emphasized in \cite{polyakov}, this action 
is invariant under scale transformations in the target space,
$ X^i \to \lambda X^i$, which is consistent with what we
expect in a conformal theory. 
Actually, it is possible to prove that the action is also invariant
under special conformal transformations, so that it is invariant under
the full conformal group\footnote{It is enough to prove that the action
is invariant under inversions $X^i \rightarrow X^i/X^2$. This can be
shown using identities like $\nabla_\beta X^i \nabla^2 X^i =0$, which 
use the fact that $\gamma_{\alpha \beta} = \partial_\alpha
 X^i \partial_\beta X^i$ is the induced metric.}. 
One implication of this logarithmic term is that the expectation value
of the Wilson surface is not well defined, since we can add any constant
to the logarithmic subtraction. 
Furthermore, it seems to indicate that the expectation value of a Wilson
surface is scale dependent.

It seems natural to speculate that tensionless strings in this
six-dimensional field theory are governed by some supersymmetric
form of the the action~\eqref{eq:rigid}. 

Despite the fact that the expectation value of the spherical Wilson
surface is not well defined, the connected correlation functions of
Wilson surfaces do not receive extra divergent contributions. These
correlators can be calculated in a completely analogous fashion to  
the Wilson loops in section~\ref{sec:lightscalars}. One considers a
Wilson surface whose characteristic size is much smaller than its
distance from any probe in the theory. Then one
identifies the operators that are allowed to appear in the OPE and
computes the necessary correlation functions to extract the OPE
coefficients. The details will appear in~\cite{Corrado:Correlators}. 

\section{Conclusions}

In this paper, we have made use of the the connection between
Anti--de Sitter spaces and conformal field theories in the large $N$
and $g_{YM}^2 N$ limit to compute the
operator product expansion of ``small'' Wilson loops in $\CN=4$, 
$d=4$
super Yang-Mills theory and Wilson surfaces in the $(0,2)$
superconformal theory in six dimensions.  

By determining what supergravity states couple to the worldsheet
describing the Wilson loop, we found the
set of  operators that are allowed to appear in the OPE. By
computing the amplitudes for exchange of supergravity modes between a
Wilson loop and the boundary and between two Wilson loops, we were
able to compute the correlation functions of a Wilson loop with a CFT
operator and with another loop. From these expressions, we were able
to deduce the coefficients that appear in the OPE for the states that
we considered.  

We also investigated Wilson surfaces in the (0,2) six-dimensional field theory
and we found that there is a UV
logarithmically divergent contribution to its expectation value  
 proportional to the action of a rigid string embedded
in the 5-brane worldvolume.

\section*{Acknowledgements}

We  would like to thank Philip Candelas, Jacques
Distler,  Sangmin Lee, Sasha Polyakov and Andy  Strominger for
discussions.
R.C.\ thanks the organizers of the Spring School on
Mathematics and String Theory at the Departments of Physics
and Mathematics at Harvard University and the organizers of 
{\it Strings '98} at UCSB for travel support and their kind
hospitality while some of this work was carried out.
J.M.\ and W.F.\ want to thank the Aspen Center for Physics where part
of this research was done.

\appendix

\section{Green's Functions}

Scalar Green's functions on anti-de Sitter space have been discussed
in a large number of 
papers~\cite{Witten:Holography,Gubser:Correlators,Leigh:Large-N,%
Muck:Correlators,Freedman:Correlators,Chalmers:R-current} in
connection with the types of  
correlation functions we are looking to compute. We will repeat some
of these calculations in this appendix.

Consider the action of a real scalar field $\phi$ on Euclidean anti-
de
Sitter spacetime,  of radius one, with source~$J$,
\begin{equation}
S = \int_{AdS} d^{d+1}x \, \sqrt{g}\, B  \left[ \frac{1}{2} (d\phi)^2 
+ \frac{m^2}{2} \phi^2 - \phi J \right]. \label{eq:euclidaction}
\end{equation} 
where $B$ is some constant. 
The equation of motion is
\begin{equation}
B (-\nabla^2_x + m^2)\phi=J, \label{eq:eq-of-motion}
\end{equation} 
where $\nabla^2_x= \frac{1}{\sqrt{g}}\partial_\mu(\sqrt{g}
g^{\mu\nu}\partial_\nu)$ is the Laplacian. Imposing the boundary
condition $\phi|_{\partial M} = 0$ yields a 
unique solution for $\phi$, as long as the operator $-\nabla^2_x + 
m^2$,
is positive definite. This is the case for all 
$m^2\geq -d^2/4$~\cite{Breitenlohner:PosEnergy}. 

Solutions for $\phi$ which minimize
the action~\eqref{eq:euclidaction} are given by the integral equation
\begin{equation}
\phi(x) = \int_M d^{d+1}x^\prime \sqrt{g(x^\prime)} \, G(x,x^\prime)
J(x^\prime), \label{eq:phi-sol}
\end{equation}
where the kernel $G(x,x^\prime)$ is the covariant Green's function 
for
the equation of motion~\eqref{eq:eq-of-motion}, satisfying
\begin{equation}
B (-\nabla^2_x + m^2 )G(x,x^\prime) 
= \frac{1}{\sqrt{g(x^\prime)}}\delta^{(d+1)}(x-x^\prime).
\label{eq:greens-fn} 
\end{equation}
By evaluating the action at a solution of~\eqref{eq:phi-sol}, one
finds that 
\begin{equation}
S = -\frac{1}{2} \int_M  d^{d+1}x \, \int_{M^\prime} d^{d+1}x^\prime
\, \sqrt{g(x)} \, J(x) \, G(x,x^\prime) \, \sqrt{g(x^\prime)} \,
J(x^\prime).  \label{eq:two-same-sources}
\end{equation}
If we consider the specific case of two separated sources, $J_1(x)$
and $J_2(x^\prime)$, the amplitude for their interaction is
then given by
\begin{equation}
S = -\int_M d^{d+1}x \, \int_{M^\prime}  d^{d+1}x^\prime \,
\sqrt{g(x)} \, J_1(x) \, G(x,x^\prime) \, \sqrt{g(x^\prime)} \,
J_2(x^\prime). \label{eq:twosources} 
\end{equation}

\subsection{Bulk--to--Bulk Scalar Green's Functions}

We will find it most convenient to compute the Green's function in 
the
upper-half space representation of anti-de Sitter spacetime, with
metric 
\begin{equation}
ds^2 = \frac{1}{z^2} \left( dz^2 + \sum_{i=1}^d dx_i^2 \right). 
\label{eq:UHS-metric} 
\end{equation}
As the scalar Green's function can only depend on the distance 
$d(x,x')$ 
between
the sources, in this metric it will be a function of 
\begin{equation}
W=\frac{z z^\prime}{(z - z^\prime)^2 +
\sum_{i=1}^d |x_i - x_i^\prime|^2} 
= \frac{1}{2} \, \frac{1}{\cosh d(x,x^\prime) -1}. 
\label{eq:wapprox}\end{equation}
Note that $W$ is singular precisely at $(z,x) = (z',x^\prime)$, 
which is the
location of the singularity in the Green's
function~\eqref{eq:greens-fn}.
It is easy to show that the solution for~\eqref{eq:greens-fn} which 
goes to zero at the boundary is given in terms of the hypergeometric 
function
\begin{equation}
G(W) = { \alpha_0 \over B} W^{\Delta} \, 
{_2}F_1(\Delta,\Delta + \tfrac{1-d}{2},2\Delta-d+1;-4W). 
\label{eq:UHS-Greens} 
\end{equation}
where $\Delta = d/2 + \sqrt{ m^2 + d^2/4 }$ will be the conformal weight of 
the associated operator and $\alpha_0$ is 
\begin{equation}
\begin{split} 
\alpha_0 =& \frac{
 \Gamma(\Delta)  }{
2 \pi^\frac{ d }{2} \Gamma(  \Delta -{d\over 2} +1 ) }
\\
\alpha_0 = & \frac{ \Delta -1 }{ 2 \pi^2 } ~~~~~~~~~~~~{\rm for}~~ d =4 \\
\end{split} \label{eq:alphanaught} 
\end{equation}

\subsection{Computation of correlation functions}
\label{sec:bulk-boundary}

Let us consider the field in the bulk~\eqref{eq:phi-sol} produced by 
a
source on the boundary. To be precise, we take $J(x^\prime)$ to have
support very close to the boundary, 
\begin{equation}
\text{supp}\bigl(J(x^\prime)\bigr)
= \{z^\prime | 0\leq z^\prime < \epsilon \},\
\end{equation} 
for infinitesimal $\epsilon$. Then, using the bulk-to-bulk Green's
function~\eqref{eq:UHS-Greens}, we can write 
\begin{equation}
\phi(x) = \int_{\partial M} d^{d}x^\prime \int_0^\epsilon dz^\prime
\, \sqrt{g(x^\prime)} \, G(x,x^\prime) J(x^\prime).
\end{equation}
In the region over which we integrate $z^\prime$, $W\sim 0$, so 
that
we can approximate 
\begin{equation}
G(W) \sim { \alpha_0 \over B } W^{\Delta},
\label{eq:greenapprox}\end{equation}
so that
\begin{equation}
\phi(x) \sim { \alpha_0 \over B }  \int_{\partial M} 
d^{d}x^\prime
\left( \frac{z}{z^2+|\vec{x}-\vec{x}^\prime|^2} \right)^\Delta 
\int_0^\epsilon dx_0^\prime \, {z}^{\Delta-d-1}
J(x^\prime). \label{eq:bulk-near-bound} 
\end{equation}

Now, to make contact with Witten's analysis
in~\cite{Witten:Holography}, we want to define a dimension $d-\Delta$
source $\phi_0(\vec{x}^\prime)$. 
\begin{equation}
\phi_0(\vec{x}^\prime) = c 
\int_0^\epsilon dz^\prime \, {z^\prime}^{\Delta-d-1} J(x^\prime),
\label{eq:phinaught-J}
\end{equation}
where $c$ is a numerical factor which we will determine from the
2-point function. 
Now, evaluating the action obtained from~\eqref{eq:twosources}, we
find that 
\begin{equation}
\begin{split}
S =& { \alpha_0 \over B} \int d^{d}x \, d^{d}x^\prime \,
\frac{1}{|\vec{x}-\vec{x}^\prime|^{2\Delta}} 
\int dz\, z^{\Delta-d-1} J(x)\,
\int dz^\prime \,{z^\prime}^{\Delta-d-1} J(x) \\
=& \frac{\alpha_0}{B c^2 } \int d^{d}x \, d^{d}x^\prime \,
\frac{\phi_0(\vec{x})\phi_0(\vec{x}^\prime)}{
|\vec{x}-\vec{x}^\prime|^{2\Delta}}. 
\end{split}
\end{equation}
The two-point function is then
\begin{equation}
\frac{\delta^2 S}{\delta\phi_0(\vec{x}) \delta\phi_0(\vec{x}^\prime) 
} 
= \frac{\alpha_0}{B c^2}   \,  
\frac{1}{|\vec{x}-\vec{x}^\prime|^{2\Delta}}.
\end{equation}
Choosing the convention that the operator corresponding to this 
scalar  is  unit-normalized, $\langle \CO \CO \rangle = 1/x^{2 \Delta} $,
 we determine
that $c=\sqrt{\alpha_0/B}$.

\renewcommand{\baselinestretch}{1.0} \normalsize


\bibliography{strings,m-theory,susy,largeN}
\bibliographystyle{utphys}

\end{document}